\documentclass[twocolumn,preprintnumbers,amsmath,amssymb]{revtex4}

\newcommand{\commentoutA}[1]{}

\usepackage{graphicx}
\usepackage{dcolumn}
\usepackage{bm}
\begin{document}


\title{Extended Lagrangian Born-Oppenheimer molecular dynamics
in the limit of vanishing self-consistent field optimization}

\author{Petros  Souvatzis\footnote{Email: petros.souvatsiz@fysik.uu.se}}
\affiliation{Department of Physics and Astronomy, Division of Materials Theory, Uppsala University,
Box 516, SE-75120, Uppsala, Sweden}
\author{Anders M.~N. Niklasson\footnote{Email: amn@lanl.gov}}
\affiliation{Theoretical Division, Los Alamos National Laboratory, Los Alamos, New Mexico 87545, USA}

\date{\today}

\begin{abstract}
We present an efficient general approach to first principles molecular dynamics simulations
based on extended Lagrangian Born-Oppenheimer molecular dynamics [A.M.N. Niklasson,
Phys. Rev. Lett. {\bf 100}, 123004 (2008)] in the limit of vanishing 
self-consistent field optimization. The reduction of the optimization requirement reduces 
the computational cost to a minimum, but without causing any significant loss of accuracy
or long-term energy drift. 
The optimization-free first principles molecular dynamics requires only one single diagonalization 
per time step and yields trajectories at the same level of accuracy as ``exact'', fully converged,
Born-Oppenheimer molecular dynamics simulations. The optimization-free limit 
of extended Lagrangian Born-Oppenheimer molecular dynamics therefore
represents an ideal starting point for a robust and efficient formulation
of a new generation first principles quantum mechanical molecular dynamics simulation schemes.
\end{abstract}

\maketitle

\section{Introduction}
With the rapid growth of available processing power, first principles molecular dynamics simulations, where the forces acting 
on the atoms are calculated on the fly using a quantum mechanical description of the electronic structure,
are becoming an increasingly powerful tool in materials science, chemistry and biology \cite{DMarx00}.
While some early applications where performed already four decades
ago \cite{MKarplus73,CLeforestier78}, it was not until the development of efficient
plane-wave pseudopotential methods \cite{RCar85,RemlerMadden90,DVanderbilt90,MCPayne92,GKresse93,RNBarnett93} 
based on density functional theory \cite{hohen,WKohn65} and the fast Fourier transform \cite{JWCooley65}, 
that first principles molecular dynamics simulations became broadly applicable.

There are two major approaches to first principles molecular dynamics: {\em a)} Born-Oppenheimer molecular
dynamics \cite{MKarplus73,CLeforestier78,MCPayne92,GKresse93,RNBarnett93,DMarx00} and {\em b)} extended Lagrangian Car-Parrinello
molecular dynamics \cite{RCar85,RemlerMadden90,MCPayne92,DMarx00,MTuckerman02,BHartke92,HBSchlegel01,JHerbert04,HBSchlegel01,BKirchner12,JHutter12}.
In Born-Oppenheimer molecular dynamics, the forces acting on the atoms are calculated at the relaxed
electronic ground state in each time step, which provides a well defined and often very accurate approximation.
A key problem, however, is that a straightforward implementation of Born-Oppenheimer molecular dynamics
is unstable and does not conserve energy without a high degree of convergence in the 
electronic structure calculations. If this is not achieved, the electronic system behaves like a heat sink or
source, gradually draining or adding energy to the atomic system \cite{RemlerMadden90,PPulay04}.
Several techniques have therefore been developed that attempts to improve the efficiency of
Born-Oppenheimer molecular dynamics and
reduce the computational cost of the electronic optimization procedure  \cite{TAArias92,PPulay04,ANiklasson06,TDKuhne07}.
In extended Lagrangian Car-Parrinello molecular dynamics, on the other hand, the computationally expensive ground state optimization is 
avoided.  As in Ehrenfest based molecular dynamics \cite{PEhrenfest27,JAlonso08,JJakowski09}, 
the electrons are instead treated as separate dynamical variables oscillating around 
the ground state. This approach permits a stable dynamics with a low computational cost per time step. 
Unfortunately, Car-Parrinello molecular dynamics simulations typically require shorter integration time steps and 
a system-dependent choice of electron mass parameters to yield reliable results in comparison
on an ``exact'' Born-Oppnheimer molecular dynamics; although statistical averages are often in 
good agreement \cite{DMarx00,JHutter12}.

Recently, an extended Lagrangian formulation for a time-reversible Born-Oppenheimer molecular dynamics 
was proposed \cite{ANiklasson08,PSteneteg10}, which combines some of the best features of Car-Parrinello and 
regular Born-Oppenheimer molecular dynamics, while avoiding some of their most serious shortcomings. It has been 
argued that extended Lagrangian Born-Oppenheimer molecular dynamics can be seen as a general framework
both for Born-Oppenheimer and Car-Parrinello molecular dynamics \cite{JHutter12}.
In this modern formalism of extended Lagrangian first principles molecular dynamics, 
Car-Parrinello molecular dynamics appears in the limit of vanishing
self-consistent field optimization \cite{JHutter12}. However, the optimization-free limit can be 
approached in different ways providing a variety of solutions. In this paper we show how extended
Lagrangian Born-Oppenheimer molecular dynamics, in the limit of vanishing self-consistent
field optimization, gives a first principles molecular dynamics at the same level of
accuracy as ``exact'' Born-Oppenheimer molecular dynamics, but without requiring
short integration time steps or a material dependent tuning of electron mass parameters
as in Car-Parrinello molecular dynamics. The instability from the systematic energy drift associated
with incomplete convergence of the electronic structure in regular Born-Oppenheimer
molecular dynamics is also avoided. Our work here represents a generalization and first principles
extension of recent work that was demonstrated for semi-empirical self-consistent-charge 
tight-binding simulations \cite{ANiklasson12}.

The ability to achieve a high degree of accuracy in the limit of vanishing self-consistent field optimization
serves two main purposes: 1) it simplifies the calculations with a reduction of 
the optimization cost to a minimum, and 2) it provides the ideal starting point for fully converged, 
i.e. ``exact'', Born-Oppenheimer molecular dynamics simulations when the requirement of accuracy is very high.
The optimization-free limit of extended Lagrangian Born-Oppenheimer molecular dynamics 
therefore represents an efficient and robust framework for a new generation of 
first principles molecular dynamics simulations.

\section{Extended Lagrangian Born-Oppenheimer molecular dynamics}

Extended Lagrangian Born-Oppenheimer molecular dynamics \cite{ANiklasson08} can be  formulated in terms of a Lagrangian,
\begin{equation}\begin{array}{l}
{\displaystyle
{\cal L}^{\rm XBO}({\bf R},{\bf \dot R},P_0, {\dot P_0}) = \frac{1}{2}\sum_IM_I{\dot R}_I^2  - U({\bf R}; D)  }\\
{\displaystyle  ~~~~~~ + \frac{1}{2}\mu Tr[{\dot P_0}^2] - \frac{1}{2}\mu \omega^2 Tr[(D-P_0)^2]},
\end{array}
\end{equation}
where the regular Born-Oppenheimer Lagrangian defined at the electronic ground state density matrix $D$ for a given
set of nuclear coordinated, $\{R_I\} = {\bf R}$, has been extended
with auxiliary dynamical variables for the electronic degrees of freedom, $P_0$ and ${\dot P}_0$, that
evolve in a harmonic well centered around $D$. The potential energy $U({\bf R}; D)$ is here the Hartree-Fock 
or Kohn-Sham energy functional including the ion-ion repulsion energy \cite{RMcWeeny60}. The parameter $\mu$ is a fictitious
electron mass and $\omega$ is the frequency determining the curvature of the harmonic well.
Euler-Lagrange equations, in the limit $\mu \rightarrow 0$ \cite{ANiklasson08}, gives the decoupled equations of motion:
\begin{equation}\label{EL_XLBOMD}\begin{array}{l}
{\displaystyle
M_I{\ddot R}_I = -\left.{\frac{\partial U({\bf R}; D)}{\partial R_I}}\right\rvert_{P_0}}\\
~~\\
{\displaystyle
{\ddot P}_0 = \omega^2(D-P_0)}.\\
\end{array}
\end{equation}
The partial derivative of $U$ for the nuclear coordinate $R_I$ is taken with 
respect to a constant $P_0$, since $P_0$ is an independent dynamical variable.
The equations of motion can be integrated using a time-reversible symplectic scheme, both for the nuclear
and electronic degrees of freedom \cite{AOdell09,BLeimkuhler04}. By using a time-reversible 
$P_0$ as the initial guess of the iterative self-consistent 
field (SCF) optimization procedure, 
\begin{equation}
P_0 \rightarrow P_1 \rightarrow \ldots \rightarrow P_{\infty}={\rm SCF}(P_0), 
\end{equation}
where
\begin{equation}\label{DofP}
D = \lim_{n \rightarrow \infty} D(P_n) = \lim_{n \rightarrow \infty} Z\theta\left(\mu_0I-Z^TH(P_n)Z\right)Z^T,
\end{equation}
the total Born-Oppenheimer energy,
\begin{equation}\label{Etot}
E^{\rm BO}_{\rm tot} = \frac{1}{2}\sum_IM_I{\dot R}_I^2 + U({\bf R}; D),
\end{equation}
is stable without any long-term energy drift, even in the case of approximate convergence of $P_n$  
\cite{ANiklasson08,PSteneteg10,MJCawkwell12,ANiklasson12}.
The ground state density matrix $D$ in Eq.\ (\ref{DofP}) is given from the Heaviside step function, $\theta$, 
of the converged Fockian or Kohn-Sham Hamiltonian, i.\ e.\ for $\lim_{n \rightarrow \infty} H(P_n)$, 
in an orthogonal representation, $Z^THZ$, with the step formed at the chemical potential, $\mu_0$, 
separating the occupied from the unoccupied states. The congruence transformation matrix $Z$ is given 
from the inverse Cholesky or L\"{o}wdin factorization of the overlap matrix, $S$, determined by $ZSZ^T=I$.

\subsection{Fast quantum mechanical molecular dynamics}

As $n \rightarrow 0$ in Eq.\ (\ref{DofP}), i.e. in the limit of vanishing self-consistent field optimization, 
the equations of motion for the extended Lagrangian
formulation of Born-Oppenheimer molecular dynamics, Eq.\ (\ref{EL_XLBOMD}), are given by
\begin{equation}\label{FFP_MD}\begin{array}{l}
{\displaystyle
M_I{\ddot R}_I = -\left.{\frac{\partial U({\bf R}; D(P_0))}{\partial R_I}}\right\rvert_{P_0},}\\
~~\\
{\displaystyle
{\ddot P}_0 = \omega^2(D(P_0)-P_0)}.\\
\end{array}
\end{equation}
By avoiding the self-consistent-field optimization of $P_0$, these equations of motion require only one single diagonalization
per time step in the construction of $D(P_0)$ and therefore provide a computationally fast method 
for first principles quantum mechanical molecular dynamics (fast-QMMD) \cite{ANiklasson12}. 
An alternative derivation of the fast dynamics represented by Eq.\ (\ref{FFP_MD}) that is motivated through 
a different set of arguments is given in Ref. \cite{ANiklasson12}.

To guarantee stability in the integration of the electronic degrees of freedom in Eq.\ (\ref{FFP_MD}),
using an integration time step of $\delta t$, the dimensionless integration parameter $\delta t^2 \omega^2$, 
typically needs to be rescaled by a factor $c \in [0,1]$ \cite{ANiklasson12} compared to the original
integration of extended Lagrangian Born-Oppenheimer molecular dynamics \cite{ANiklasson09}. This stability condition further
assumes convexity of the total energy functional between $P_0$ and $D(P_0)$ \cite{ANiklasson12}.

The definition of $D \equiv \lim_{n \rightarrow \infty} D(P_n)$ in Eq.\ (\ref{DofP}) and our particular choice of sequence of 
limits both for $\mu \rightarrow 0$ and $n \rightarrow 0$ are important.
For example, if we instead use $D \equiv P_n$ and let $n \rightarrow 0$ in 
the Lagrangian (before deriving the Euler-Lagrange equations of motion), 
we end up with a $\mu$-dependent set of unconstrained Car-Parrinello-like equations \cite{JHutter12}
and if $\mu \rightarrow 0$ already in the initial Lagrangian, but with full self-consistency convergence, 
we recover (trivially) regular Born-Oppenheimer molecular dynamics.
For our particular sequence of limits of $\mu$ and $n$, the fast-QMMD defined by Eq.\ (\ref{FFP_MD}) is formally neither 
an extended Lagrangian nor a Born-Oppenheimer molecular dynamics.  However, as will be demonstrated in our examples, 
the first principles fast-QMMD in Eq.\ (\ref{FFP_MD}) is a very close approximation of ``exact'', 
fully converged, extended Lagrangian Born-Oppenheimer molecular dynamics.

\section{Examples}

\subsection{Implementation}

Our fast-QMMD, Eq.\ (\ref{FFP_MD}), has been implemented based on Hartree-Fock theory in the Uppsala Quantum Chemistry (UQuantChem) 
simulations package \cite{UQuantChem}, which is a freely available suite of programs for parallel {\em ab initio} electronic
structure calculations using Gaussian basis sets, including Hartree-Fock and M{\o}ller-Plesset perturbation theory, 
configuration interaction, variational and diffusion Monte-Carlo, structural optimization, and first principles 
molecular dynamics. The nuclear coordinates are integrated using the velocity Verlet scheme and the electronic
degrees of freedom with a modified Verlet algorithm, including a weak dissipation term to remove
the accumulation of numerical noise \cite{ANiklasson09,PSteneteg10}.
Since $P_0$ appears as a dynamical variable in Eq.\ (\ref{FFP_MD}), a Hellmann-Feynman-like
expression for the nuclear forces, under the constraint of $P_0$ being constant, can be applied. 
Thus, even if the ground state condition necessary for Hellmann-Feynman forces are not fulfilled,
we still have a force expression of similar simplicity.
For the basis-set dependent contribution we use the original expression of the Pulay force term \cite{PPulay69},
which provides a sufficiently accurate approximation \cite{ANiklasson12}.
Our first principles dynamics is implemented based on Hartree-Fock theory \cite{Roothaan,RMcWeeny60}.
The Hartree-Fock method is the starting point for correlated wavefunction methods and can be used as the computational
prototype for density functional theory \cite{hohen,WKohn65,RParr89,RMDreizler90} and hybrid schemes \cite{ADBecke93}.
Our optimization-free Hartree-Fock molecular dynamics therefore demonstrates
applicability for a broad class of first principles methods. Extensions to plane wave schemes should 
also be straightforward \cite{PSteneteg10}.

\begin{figure}[tbp]
\includegraphics*[angle=0,scale=0.3]{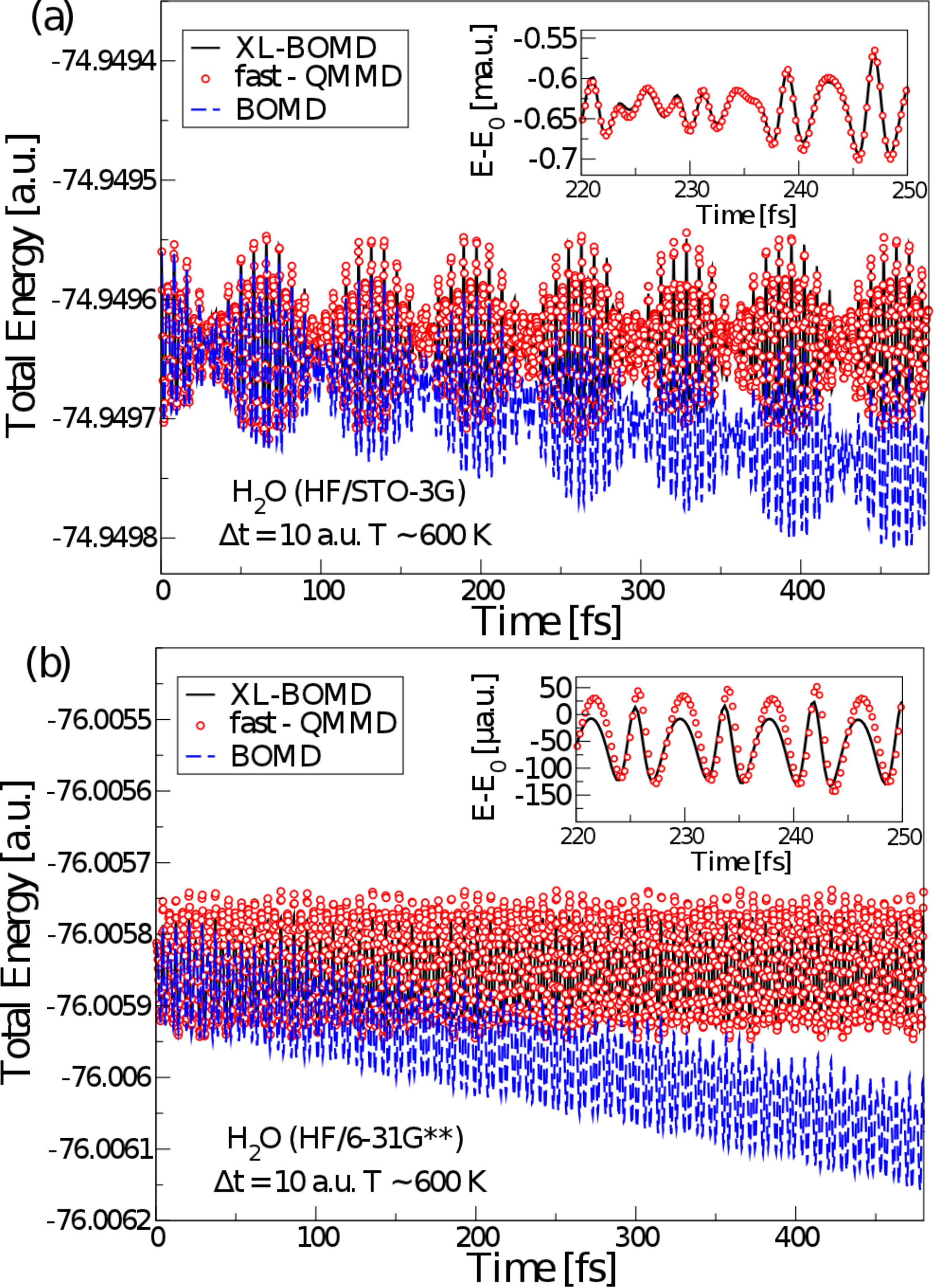}
\caption{\label{Figure_1}\small
Total energy fluctuations for water using
``exact'' (5 SCF/step) Born-Oppenheimer molecular dynamics (XL-BOMD), Eq.\ (\ref{EL_XLBOMD}),
and the first principles fast-QMMD, i.e. XL-BOMD in the limit $n \rightarrow 0$, Eq.\ (\ref{FFP_MD}), 
in comparison to regular Born-Oppenheimer molecular dynamics (BOMD), where the density matrix
form the previous time step is used as the initial guess to the SCF optimization with
the energy converged to $< 0.01~ \mu$Hartree. In (a) a STO-3G basis set was used, in the inset  $E_{0}$ = -74.949 au. 
In (b) a 6-31G$^{**}$ basis set was used, in the inset $E_{0}$ = -76.0058 au}
\end{figure}

\begin{figure}[tbp]
\includegraphics*[angle=0,scale=0.3]{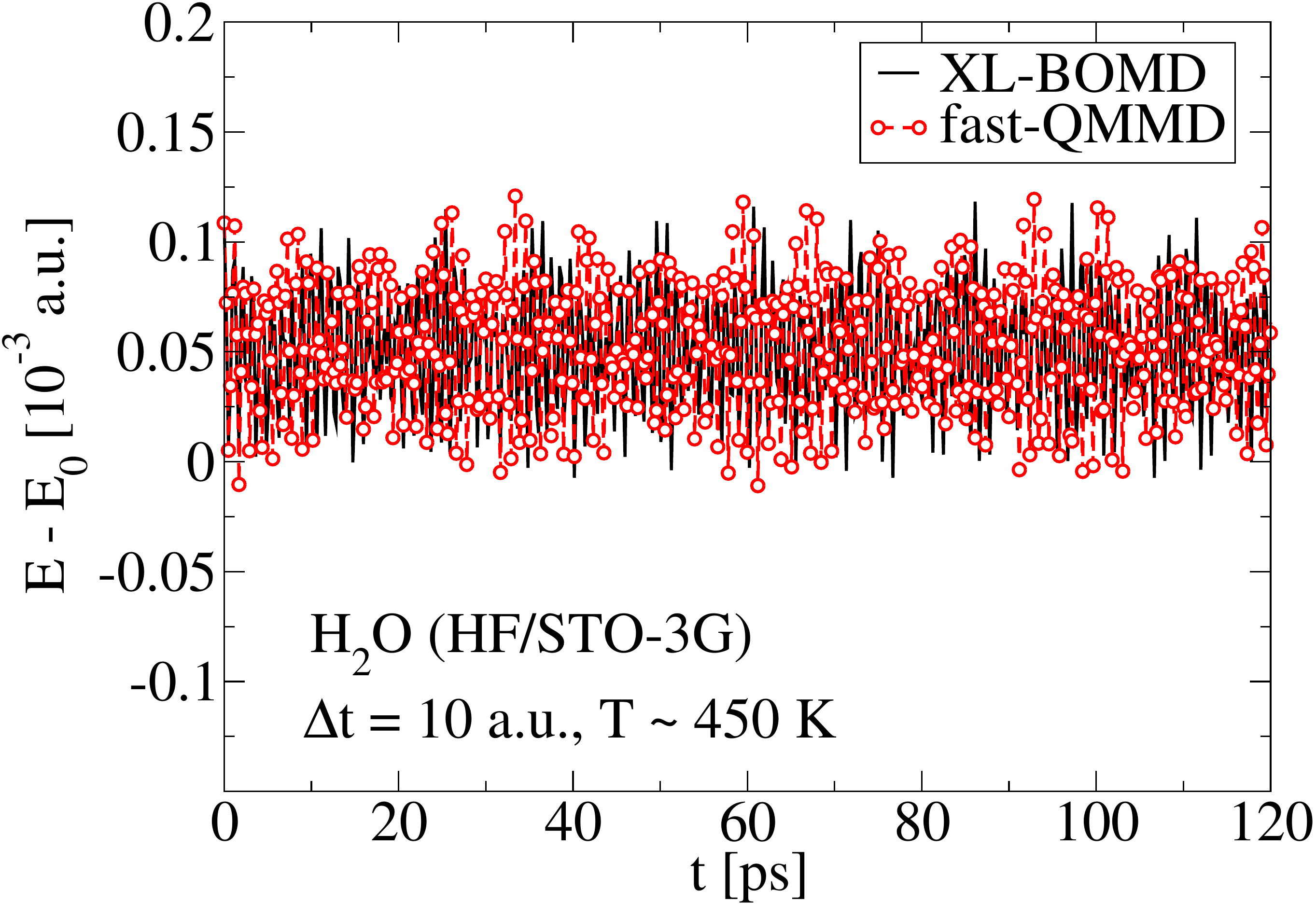}
\caption{\label{Figure_2}\small
Total energy fluctuations for a 130 ps simulation, using
``exact'' (5 SCF/step) Born-Oppenheimer molecular dynamics (XL-BOMD), Eq.\ (\ref{EL_XLBOMD}),
and the fast-QMMD, Eq.\ (\ref{FFP_MD}). Here $E_{0}$ = -74.953 a.u.}
\end{figure}

\begin{figure}[tbp]
\includegraphics*[angle=0,scale=0.37]{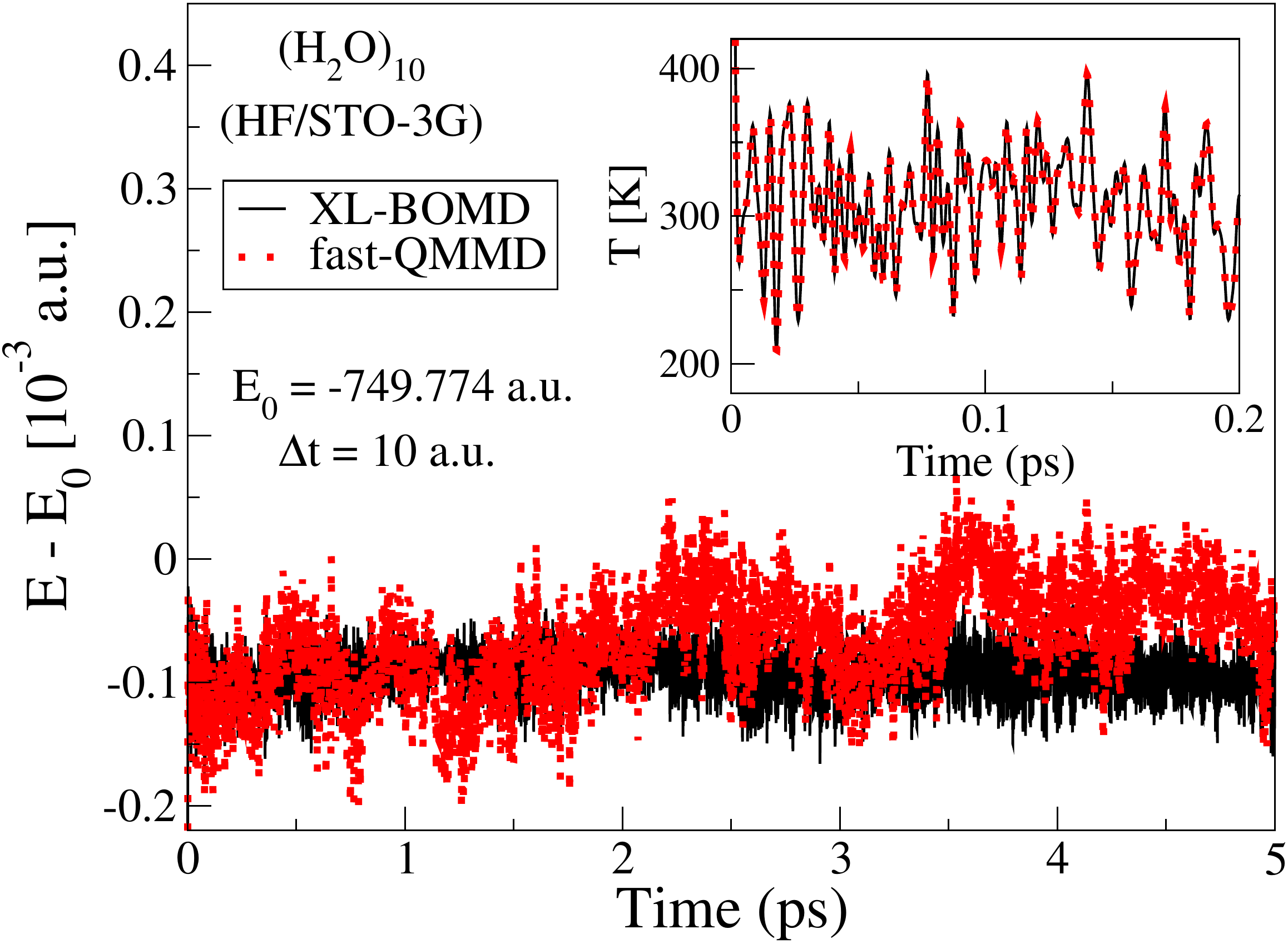}
\caption{\label{Figure_2b}\small
Total energy fluctuations for a 5 ps simulation of a small cluster containing 10 water molecules, using
``exact'' (5 SCF/step) Born-Oppenheimer molecular dynamics (XL-BOMD), Eq.\ (\ref{EL_XLBOMD}),
and the fast-QMMD, Eq.\ (\ref{FFP_MD}). Here $E_{0}$ = -749.774 a.u. The inset shows the temperature fluctuations
for the first 200 fs of simulation.}
\end{figure}

\begin{figure}[tbp]
\includegraphics*[angle=0,scale=0.33]{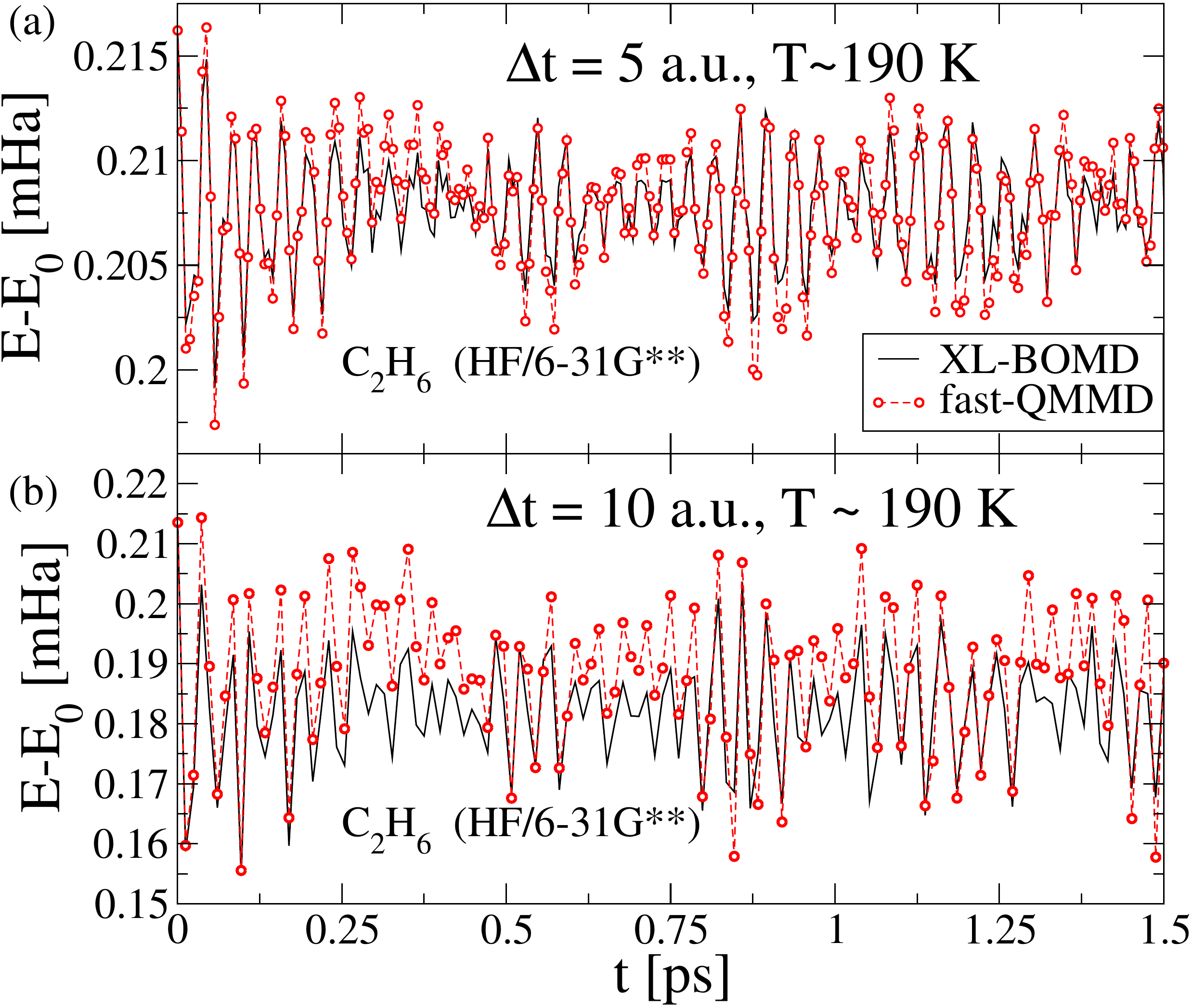}
\caption{\label{Figure_3}\small
Total energy fluctuations for ethane using
``exact'' (5 SCF/step) Born-Oppenheimer molecular dynamics (XL-BOMD), Eq.\ (\ref{EL_XLBOMD}),
and the first principles fast-QMMD, i.e. XL-BOMD in the limit $n \rightarrow 0$, Eq.\ (\ref{FFP_MD}). 
In the upper panel (a) a time-step of $\Delta t$ = 5 au  was used.
In the lower panel (b) a time-step of $\Delta t$ = 10 au  was used. Here $E_{0}$ = -79.224 a.u. }
\end{figure}

\begin{figure}[tbp]
\includegraphics*[angle=0,scale=0.33]{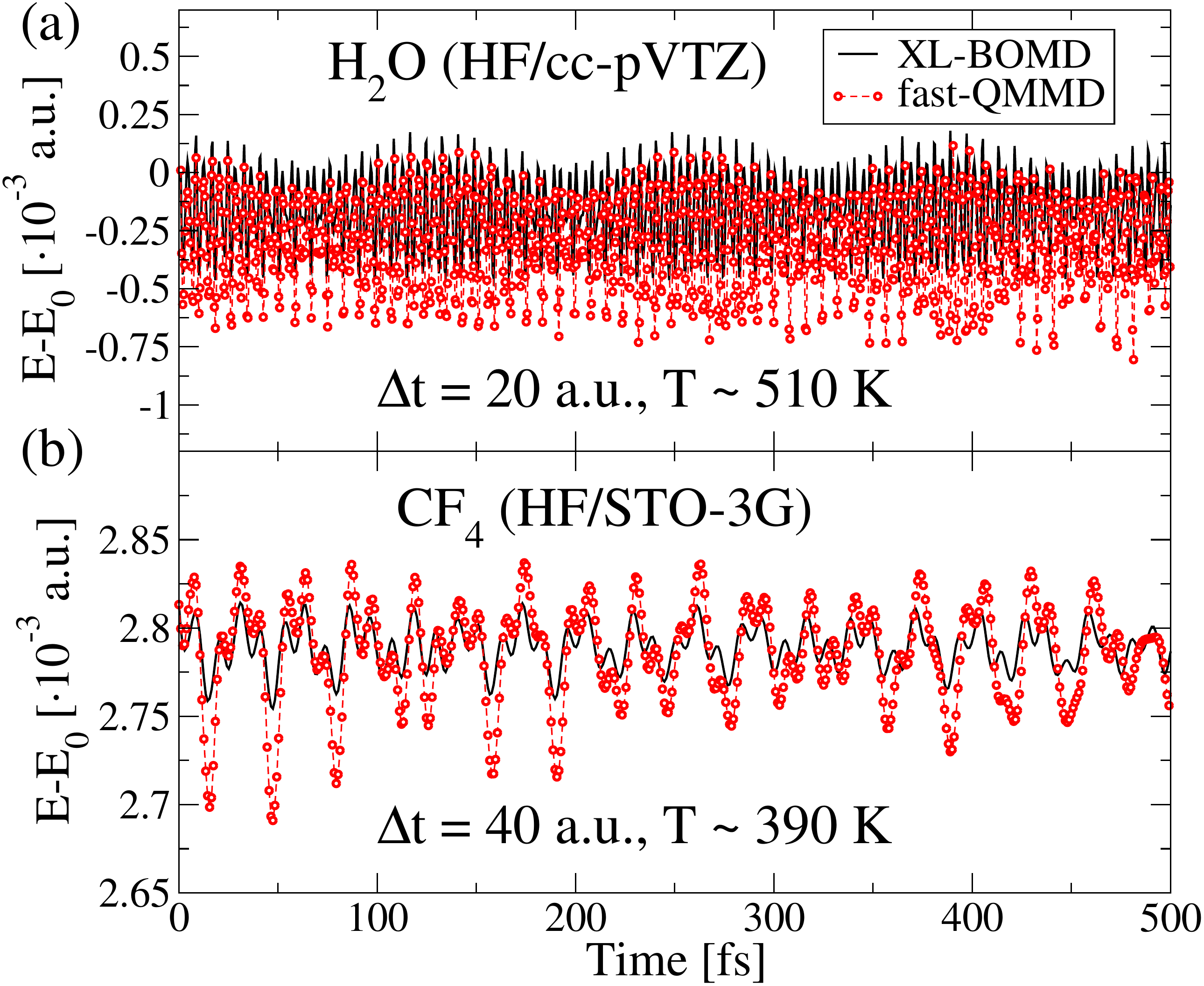}
\caption{\label{Fig_5}\small
Total energy fluctuations calculated at a level of URHF theory for a H$_{2}$O molecule, (a), and a CF$_{4}$ molecule, (b), using
``exact'' (5 SCF/step) Born-Oppenheimer molecular dynamics (XL-BOMD), Eq.\ (\ref{EL_XLBOMD}),
and the first principles fast-QMMD, i.e. XL-BOMD in the limit $n \rightarrow 0$, Eq.\ (\ref{FFP_MD}). 
In the upper panel (a) a time-step of $\Delta t$ = 20 a.u.  was used.
In the lower panel (b) a time-step of $\Delta t$ = 40 a.u.  was used. Here $E_{0}$ = -76.043 a.u., in (a), 
and $E_{0}$ = -429.57 a.u., in (b). }
\end{figure}

\begin{figure}[tbp]
\includegraphics*[angle=0,scale=0.33]{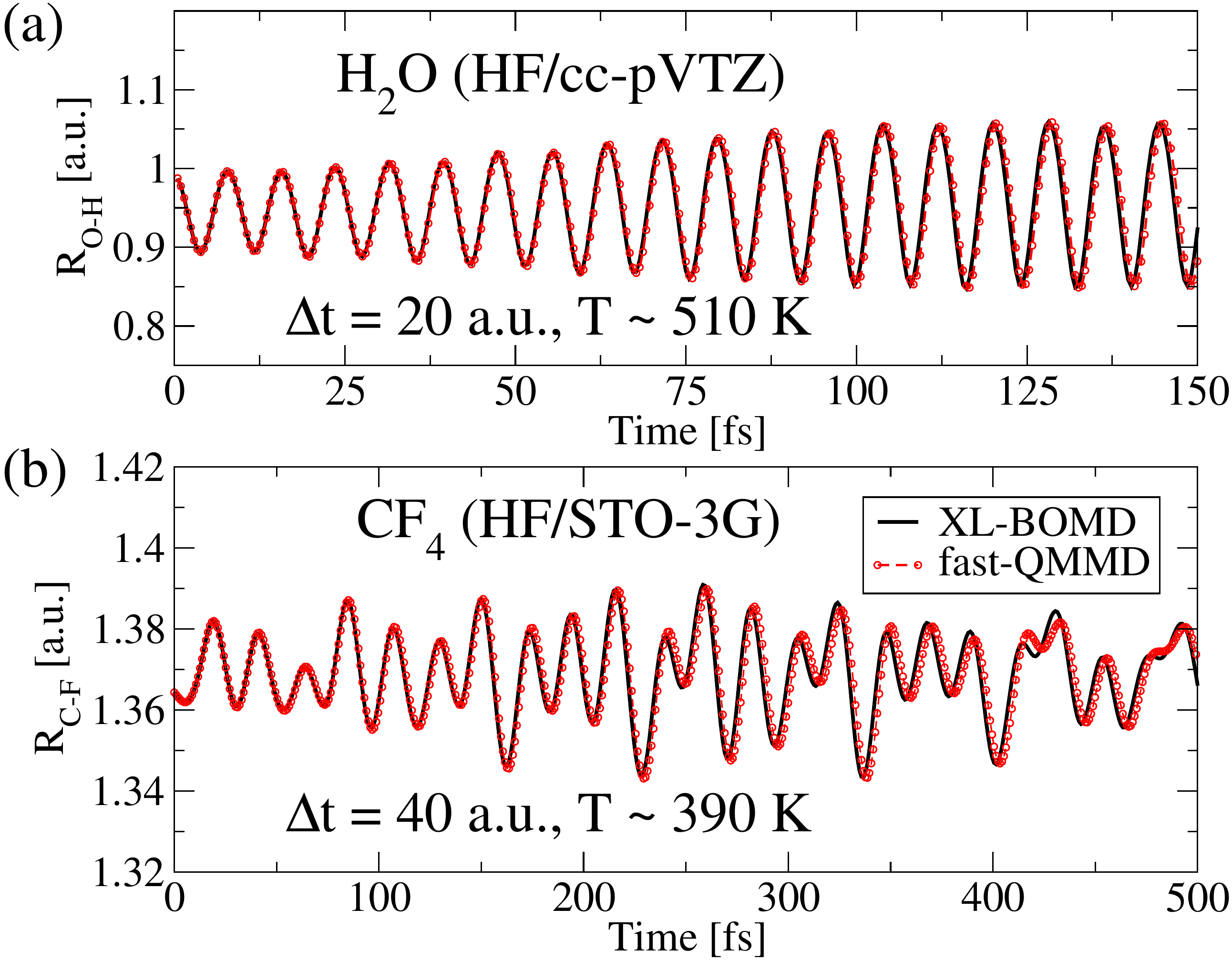}
\caption{\label{Fig_6}\small
Interatomic distances calculated at a level of URHF theory, using
``exact'' (5 SCF/step) Born-Oppenheimer molecular dynamics (XL-BOMD), Eq.\ (\ref{EL_XLBOMD}),
and the first principles fast-QMMD, i.e. XL-BOMD in the limit $n \rightarrow 0$, Eq.\ (\ref{FFP_MD}).  In (a), the interatomic distance between the Oxygen atom 
and one of the the Hydrogen atoms, R$_{O-H}$, in a H$_{2}$O molecule. In (b), the interatomic distance between the Carbon atom 
and one of the Fluorine atoms, R$_{C-F}$, in a CF$_{4}$ molecule.}
\end{figure}

\begin{figure}[tbp]
\includegraphics*[angle=0,scale=0.32]{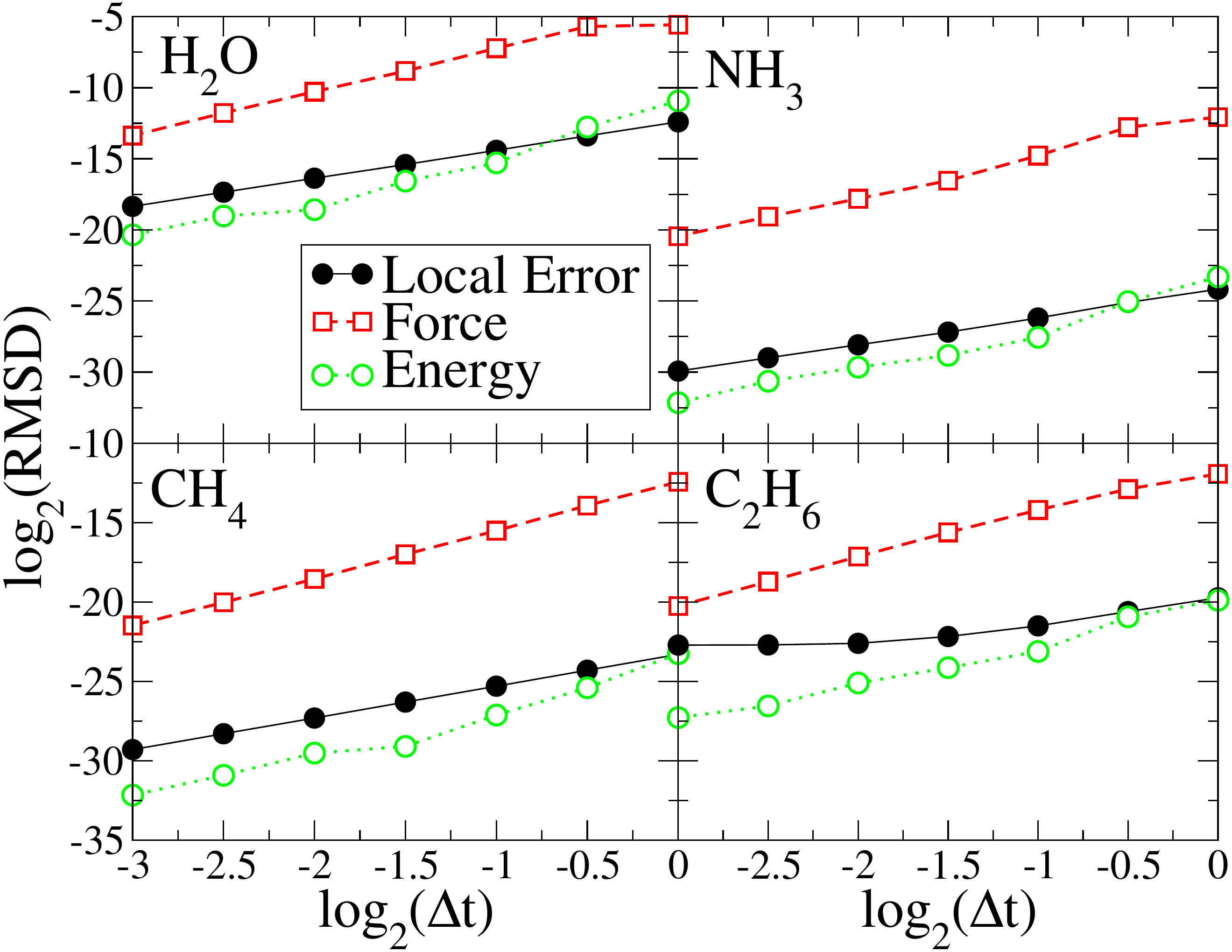}
\caption{\label{Figure_4}\small
The root mean square deviation (RMSD) between the fast quantum mechanical molecular dynamics and ``exact'' (5 SCFs/step) 
Born-Oppenheimer molecular dynamics, Eq.\ (\ref{EL_XLBOMD}), for the nuclear forces (red squares) and for the total energy (green circles)
calculated for four different molecules at different time steps. For comparison the local error of the total 
energy (black filled circles) has been calculated. Simulations were performed with a  6-31G$^{**}$ basis set, 
using URHF theory as implemented in the UQuantChem code \cite{UQuantChem}.}
\end{figure}

\begin{figure}[tbp]
\includegraphics*[angle=0,scale=0.3]{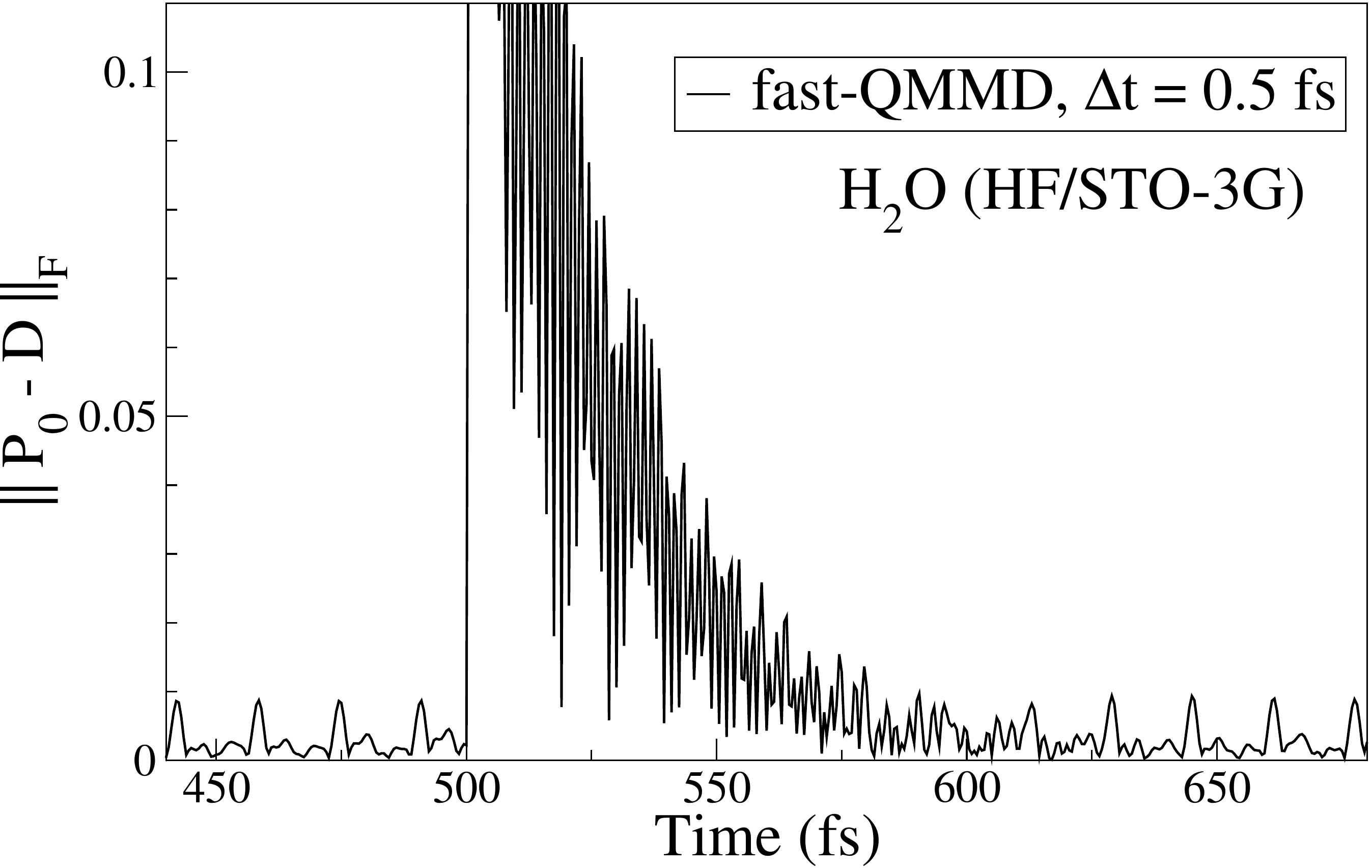}
\caption{\label{Figure_5}\small
The deviation as measured by the Frobenius norm between the fast-QMMD density matrix, $P_0$, 
and the "exact'' (5 SCF/step) Born-Oppenheimer 
molecular dynamics, Eq.\ (\ref{EL_XLBOMD}), (XL-BOMD) density matrix, $D(P_5)$, after perturbing the density matrix, 
$P_0(t)$, at $t=500$ fs, by resetting $P_0$ at $t=500$ fs to the initial density matrix at $t=0.5$ fs.  
Simulations were performed for a single water molecule at room temperature with a time step of 0.5 fs  using 
URHF theory  as implemented in the UQuantChem code \cite{UQuantChem}.}
\end{figure}

\subsection{Molecular dynamics simulations}

Figure \ref{Figure_1} shows the behavior of the total energy, Eq.\ (\ref{Etot}), for the simulation of a single
water molecule using a STO-3G basis set in (a) and a 6-31G$^{**}$ in (b). 
Regular Born-Oppenheimer molecular dynamics, where the density matrix from the previous time step
was used as the initial guess to the iterative ground-state optimization, exhibits an unphysical systematic drift 
in the total energy because of the broken time-reversal symmetry \cite{PPulay04,ANiklasson08}. This drift
is avoided in the ``exact'' fully optimized extended Lagrangian Born-Oppenheimer molecular dynamics (XL-BOMD), Eq.\ (\ref{EL_XLBOMD}),
which is very close to the results from the optimization-free fast
first principles QMMD (red circles), Eq.\ (\ref{FFP_MD}), as seen in the insets.
In particular, any deviations between the optimization-free and the fully optimized Born-Oppenheimer molecular dynamics
simulations are small compared to the local truncation error, i.e. the amplitude of the total energy fluctuations that are
caused by the finite size of the integration time step $\delta t=10$ a.\ u. As in classical molecular dynamics, the
dominating integration error is thus determined by our choice of integration scheme and the size of the time step.
Figure \ref{Figure_2} demonstrates the long-term stability of our first principles fast-QMMD, which
shows no systematic drift in the energy over 120 ps of simulation time. 
However, for longer integration time steps we have occasionally
noticed a small drift that seems to be caused by the dissipation force of the modified Verlet scheme. This sensitivity, which
not yet is completely understood, is not found in partially or fully SCF optimized versions of extended Lagrangian 
Born-Oppenheimer molecular dynamics. The next figure, Fig. \ref{Figure_2b}, shows the corresponding simulation
for a water cluster containing 10 water molecules simulated for a shorter simulation time. Because of the
chaotic movements of the larger system, a direct comparison with respect to the total energy is harder. The
inset shows a comparison of the kinetic energy fluctuations given by the temperature over the first 200 fs of simulation time, shortly
before they eventually get out of phase. The total energy fluctuations of the fast-QMMD simulation shows a 
noisy behavior similar to a random walk compared to the ``exact'' Born-Oppenheimer simulation (XL-BOMD). Similar
random walk-like noise have been seen in linear scaling XL-BOMD simulations \cite{MJCawkwell12}.

Figure \ref{Figure_3} shows the behavior of the total energy, Eq.\ (\ref{Etot}), in simulations of 
a $C_{2}H_{6}$ molecule using a 6-31G$^{**}$ basis set, which represents a slightly larger and more complex
system compared to the water molecule. As a comparison two different time steps were used, in panel (a) $\Delta t = 5$ au 
and in panel (b) $\Delta t = 10$ au.

In Figure \ref{Fig_5},  simulations of a $H_{2}O$ and a $CF_{4}$ molecule, showing the behavior of the total energy, Eq.\ (\ref{Etot}), 
using two times respectively four times as long time steps as the maximum time step used in the previous examples. 

Figure \ref{Fig_6} illustrates the behavior of the interatomic distances, in simulations of a $H_{2}O$ and a $CF_{4}$ molecule.

Figure \ref{Figure_4} shows the convergence toward ``exact'' Born-Oppenheimer molecular dynamics as the length of the integration
time step $\delta t$ is reduced. 
This scaling demonstrates how the fast-QMMD scheme provides a well defined approximation to 
exact Born-Oppenheimer molecular dynamics whith an error of order $\delta t^2$, i.e.
\begin{equation}\label{FFP_MD_ERR}\begin{array}{l}
{\displaystyle
M_I{\ddot R}_I  = -\left.{\frac{\partial U({\bf R}; D(P_0))}{\partial R_I}}\right\rvert_{P_0}} + {\cal O}(\delta t^2)\\
~~\\
{\displaystyle
{\ddot P}_0 = \omega^2(D(P_0)-P_0)} + {\cal O}(\delta t^2).\\
\end{array}
\end{equation}
The corresponding behavior was recently found in our studies based on self-consistent-charge 
tight-binding simulations \cite{ANiklasson12}.

To illustrate the stability of first principles fast-QMMD we perturb a simulation by resetting
the auxiliary density matrix $P_0(t)$ to its $t_0$ initial value after 500 fs of simulation time.
During the continued simulation, the perturbation slowly disappears as $P_0(t)$ converges toward the electronic ground state, as seen  in Figure \ref{Figure_5}, where the deviation of the fast-QMMD density matrix $P_{0}$
relative to the ``exact'' density matrix $P_{5}$ is plotted as a function of time.
This behavior demonstrates a key mechanism of our method. Instead 
of optimizing to ground state in each iteration as in regular Born-Oppenheimer molecular 
dynamics, the {\em time evolution} of the electronic degrees of freedom makes $P_0(t)$ 
converge toward the ground state dynamically. At convergence, the auxiliary density matrix
$P_0(t)$ oscillates around the exact ground state with an amplitude that is of the order $\delta t^{2}$.

\section{Conclusions and summary}

The extended Lagrangian approach to first principles molecular dynamics, as pioneered by Roberto Car 
and Michele Parrinello \cite{RCar85}, in its modern formulation of extended Lagrangian Born-Oppenheimer molecular 
dynamics \cite{ANiklasson08,JHutter12}, provides an efficient and versatile framework for first principles molecular dynamics simulations.
Here we have shown how the ground state optimization requirement can be simplified and reduced to a minimum without 
causing any significant loss of accuracy or long-term stability. This has been demonstrated using 
Hartree-Fock theory and should be applicable to a broad class of first principles methods.
The optimization-free first principles molecular dynamics requires only one single diagonalization
per time step and yields trajectories that are very close to an ``exact'', time-reversible, 
first principles Born-Oppenheimer molecular dynamics simulation. 

\section{Acknowledgements}

P. S. wants to thank L. S. for her eternal patience. 
A.M.N.N acknowledge support by the United States Department of Energy (U.S. DOE) Office
of Basic Energy Sciences as well as discussisions with C.J. Tymczak and stimulating contributions 
by T. Peery at the T-Division Ten Bar Java group.  LANL is operated by Los Alamos National Security, LLC,
for the NNSA of the U.S. DOE under Contract No. DE-AC52- 06NA25396.


\begin{thebibliography}{29}
\expandafter\ifx\csname natexlab\endcsname\relax\def\natexlab#1{#1}\fi
\expandafter\ifx\csname bibnamefont\endcsname\relax
  \def\bibnamefont#1{#1}\fi
\expandafter\ifx\csname bibfnamefont\endcsname\relax
  \def\bibfnamefont#1{#1}\fi
\expandafter\ifx\csname citenamefont\endcsname\relax
  \def\citenamefont#1{#1}\fi
\expandafter\ifx\csname url\endcsname\relax
  \def\url#1{\texttt{#1}}\fi
\expandafter\ifx\csname urlprefix\endcsname\relax\def\urlprefix{URL }\fi
\providecommand{\bibinfo}[2]{#2}
\providecommand{\eprint}[2][]{\url{#2}}

\bibitem[{\citenamefont{Marx and Hutter}(2000)}]{DMarx00}
\bibinfo{author}{\bibfnamefont{D.}~\bibnamefont{Marx}} \bibnamefont{and}
  \bibinfo{author}{\bibfnamefont{J.}~\bibnamefont{Hutter}},
  \emph{\bibinfo{title}{Modern Methods and Algorithms of Quantum Chemistry}}
  (\bibinfo{publisher}{ed. J. Grotendorst}, \bibinfo{address}{John von Neumann
  Institute for Computing, J\"ulich, Germany}, \bibinfo{year}{2000}),
  \bibinfo{edition}{2nd} ed.

\bibitem[{\citenamefont{Wang and Karplus}(1973)}]{MKarplus73}
\bibinfo{author}{\bibfnamefont{I.~S.~Y.} \bibnamefont{Wang}} \bibnamefont{and}
  \bibinfo{author}{\bibfnamefont{M.}~\bibnamefont{Karplus}},
  \bibinfo{journal}{J. Am. Chem. Soc.} \textbf{\bibinfo{volume}{95}},
  \bibinfo{pages}{8160} (\bibinfo{year}{1973}).

\bibitem[{\citenamefont{Leforestier}(1978)}]{CLeforestier78}
\bibinfo{author}{\bibfnamefont{C.}~\bibnamefont{Leforestier}},
  \bibinfo{journal}{J. Chem. Phys.} \textbf{\bibinfo{volume}{68}},
  \bibinfo{pages}{4406} (\bibinfo{year}{1978}).

\bibitem[{\citenamefont{Car and Parrinello}(1985)}]{RCar85}
\bibinfo{author}{\bibfnamefont{R.}~\bibnamefont{Car}} \bibnamefont{and}
  \bibinfo{author}{\bibfnamefont{M.}~\bibnamefont{Parrinello}},
  \bibinfo{journal}{Phys.\ Rev.\ Lett.} \textbf{\bibinfo{volume}{55}},
  \bibinfo{pages}{2471} (\bibinfo{year}{1985}).

\bibitem[{\citenamefont{Remler and Madden}(1965)}]{RemlerMadden90}
\bibinfo{author}{\bibfnamefont{D.~K.}~\bibnamefont{Remler}} \bibnamefont{and}
  \bibinfo{author}{\bibfnamefont{P.~A.} \bibnamefont{Madden}},
  \bibinfo{journal}{Mol. Phys.} \textbf{\bibinfo{volume}{70}},
  \bibinfo{pages}{921} (\bibinfo{year}{1990}).

\bibitem[{\citenamefont{Vanderbilt}(1990)}]{DVanderbilt90}
\bibinfo{author}{\bibfnamefont{D.}~\bibnamefont{Vanderbilt}},
  \bibinfo{journal}{Phys.\ Rev.\ B} \textbf{\bibinfo{volume}{41}},
  \bibinfo{pages}{7892} (\bibinfo{year}{1990}).

\bibitem[{\citenamefont{Payne et~al.}(1992)\citenamefont{Payne, Teter, Allan,
  Arias, and Joannopoulos}}]{MCPayne92}
\bibinfo{author}{\bibfnamefont{M.~C.} \bibnamefont{Payne}},
  \bibinfo{author}{\bibfnamefont{M.~P.} \bibnamefont{Teter}},
  \bibinfo{author}{\bibfnamefont{D.~C.} \bibnamefont{Allan}},
  \bibinfo{author}{\bibfnamefont{T.~A.} \bibnamefont{Arias}}, \bibnamefont{and}
  \bibinfo{author}{\bibfnamefont{J.~D.} \bibnamefont{Joannopoulos}},
  \bibinfo{journal}{Rev. Mod. Phys.} \textbf{\bibinfo{volume}{64}},
  \bibinfo{pages}{1045} (\bibinfo{year}{1992}).

\bibitem[{\citenamefont{Kresse}(1993)}]{GKresse93}
\bibinfo{author}{\bibfnamefont{G.} \bibnamefont{Kresse}}, \bibnamefont{and}
\bibinfo{author}{\bibfnamefont{J.} \bibnamefont{Hafner}},
  \bibinfo{journal}{Phys.\ Rev.\ B} \textbf{\bibinfo{volume}{47}},
  \bibinfo{pages}{558} (\bibinfo{year}{1993}).

\bibitem[{\citenamefont{Barnett}(1993)}]{RNBarnett93}
\bibinfo{author}{\bibfnamefont{R.~N.} \bibnamefont{Barnett}}, \bibnamefont{and}
\bibinfo{author}{\bibfnamefont{U.} \bibnamefont{Landman}},
  \bibinfo{journal}{Phys.\ Rev.\ B} \textbf{\bibinfo{volume}{48}},
  \bibinfo{pages}{2081} (\bibinfo{year}{1993}).

\bibitem[{\citenamefont{Hohenberg and Kohn}(1964)}]{hohen}
\bibinfo{author}{\bibfnamefont{P.}~\bibnamefont{Hohenberg}} \bibnamefont{and}
  \bibinfo{author}{\bibfnamefont{W.}~\bibnamefont{Kohn}},
  \bibinfo{journal}{Phys. Rev.} \textbf{\bibinfo{volume}{136}},
  \bibinfo{pages}{{\rm B}:864} (\bibinfo{year}{1964}).

\bibitem[{\citenamefont{Kohn and Sham}(1965)}]{WKohn65}
\bibinfo{author}{\bibfnamefont{W.}~\bibnamefont{Kohn}} \bibnamefont{and}
  \bibinfo{author}{\bibfnamefont{L.~J.} \bibnamefont{Sham}},
  \bibinfo{journal}{Phys. Rev. B} \textbf{\bibinfo{volume}{140}},
  \bibinfo{pages}{A1133} (\bibinfo{year}{1965}).

\bibitem[{\citenamefont{Cooley and Tukey}(1965)}]{JWCooley65}
\bibinfo{author}{\bibfnamefont{J.~W.}~\bibnamefont{Cooley}} \bibnamefont{and}
  \bibinfo{author}{\bibfnamefont{J.~W.} \bibnamefont{Tukey}},
  \bibinfo{journal}{Math. Comp.} \textbf{\bibinfo{volume}{19}},
  \bibinfo{pages}{297} (\bibinfo{year}{1965}).


\bibitem[{\citenamefont{Tuckerman}(2002)}]{MTuckerman02}
\bibinfo{author}{\bibfnamefont{M.}~\bibnamefont{Tuckerman}},
  \bibinfo{journal}{J. Phys.:Condens. Matter} \textbf{\bibinfo{volume}{50}},
  \bibinfo{pages}{1297} (\bibinfo{year}{2002}).

\bibitem[{\citenamefont{Hartke}(1992)}]{BHartke92}
\bibinfo{author}{\bibfnamefont{B.} \bibnamefont{Hartke}}, \bibnamefont{and}
\bibinfo{author}{\bibfnamefont{E.~ A.} \bibnamefont{Carter}},
  \bibinfo{journal}{Chem. Phys. Lett.} \textbf{\bibinfo{volume}{189}},
  \bibinfo{pages}{358} (\bibinfo{year}{1992}).

\bibitem[{\citenamefont{Schlegel}(2001)}]{HBSchlegel01}
\bibinfo{author}{\bibfnamefont{H.~ B.} \bibnamefont{Schlegel}}, 
\bibinfo{author}{\bibfnamefont{J.~ M.} \bibnamefont{Millam}}, 
\bibinfo{author}{\bibfnamefont{S.~ S.} \bibnamefont{Iyengar}}, 
\bibinfo{author}{\bibfnamefont{G.~ A.} \bibnamefont{Voth}}, 
\bibinfo{author}{\bibfnamefont{A.~ D.} \bibnamefont{Daniels}},
\bibinfo{author}{\bibfnamefont{G.} \bibnamefont{Scusseria}}, \bibnamefont{and}
\bibinfo{author}{\bibfnamefont{M.~ J.} \bibnamefont{Frisch}},
  \bibinfo{journal}{J. Chem. Phys.} \textbf{\bibinfo{volume}{114}},
  \bibinfo{pages}{9758} (\bibinfo{year}{2001}).

\bibitem[{\citenamefont{Herbert}(1992)}]{JHerbert04}
\bibinfo{author}{\bibfnamefont{J.} \bibnamefont{Herbert}},  \bibnamefont{and}
\bibinfo{author}{\bibfnamefont{M.} \bibnamefont{Head-Gordon}},
  \bibinfo{journal}{J. Chem. Phys.} \textbf{\bibinfo{volume}{121}},
  \bibinfo{pages}{11542} (\bibinfo{year}{2004}).

\bibitem[{\citenamefont{Schlegel}(2001)}]{HBSchlegel01}
\bibinfo{author}{\bibfnamefont{H.~ B.} \bibnamefont{Schlegel}},
\bibinfo{author}{\bibfnamefont{S.} \bibnamefont{Srinivasan}},
\bibinfo{author}{\bibfnamefont{S.~ S.} \bibnamefont{Iyengar}},
\bibinfo{author}{\bibfnamefont{X.} \bibnamefont{Li}},
\bibinfo{author}{\bibfnamefont{J.~ M.} \bibnamefont{Millam}},
\bibinfo{author}{\bibfnamefont{G.~ A.} \bibnamefont{Voth}},
\bibinfo{author}{\bibfnamefont{G.} \bibnamefont{Scusseria}}, \bibnamefont{and}
\bibinfo{author}{\bibfnamefont{M.~ J.} \bibnamefont{Frisch}},
  \bibinfo{journal}{J. Chem. Phys.} \textbf{\bibinfo{volume}{117}},
  \bibinfo{pages}{8694} (\bibinfo{year}{2002}).

\bibitem[{\citenamefont{Kirchner}(2012)}]{BKirchner12}
\bibinfo{author}{\bibfnamefont{B.} \bibnamefont{Kirchner}} 
  \bibinfo{author}{\bibfnamefont{J.~di~Dio} \bibnamefont{Philipp}}, \bibnamefont{and}
  \bibinfo{author}{\bibfnamefont{J.} \bibnamefont{Hutter}}, 
  \bibinfo{journal}{Top. Curr. Chem.} \textbf{\bibinfo{volume}{307}},
  \bibinfo{pages}{109}
  \bibinfo{publisher}{Springer Verlag}, \bibinfo{address}{Berlin Heidelberg},
  (\bibinfo{year}{2012}).

\bibitem[{\citenamefont{Hutter}(2012)}]{JHutter12}
\bibinfo{author}{\bibfnamefont{J.}~\bibnamefont{Hutter}},
  \bibinfo{journal}{WIREs Comput. Mol. Sci.} \textbf{\bibinfo{volume}{2}},
  \bibinfo{pages}{604} (\bibinfo{year}{2012}).


\bibitem[{\citenamefont{Pulay and Fogarasi}(2004)}]{PPulay04}
\bibinfo{author}{\bibfnamefont{P.}~\bibnamefont{Pulay}} \bibnamefont{and}
  \bibinfo{author}{\bibfnamefont{G.}~\bibnamefont{Fogarasi}},
  \bibinfo{journal}{Chem. Phys. Lett.} \textbf{\bibinfo{volume}{386}},
  \bibinfo{pages}{272} (\bibinfo{year}{2004}).

\bibitem[{\citenamefont{Arias et~al.}(1992)\citenamefont{Arias, Payne, and
  Joannopoulos}}]{TAArias92}
\bibinfo{author}{\bibfnamefont{T.}~\bibnamefont{Arias}},
  \bibinfo{author}{\bibfnamefont{M.}~\bibnamefont{Payne}}, \bibnamefont{and}
  \bibinfo{author}{\bibfnamefont{J.}~\bibnamefont{Joannopoulos}},
  \bibinfo{journal}{Phys. Rev. Lett.} \textbf{\bibinfo{volume}{69}},
  \bibinfo{pages}{1077} (\bibinfo{year}{1992}).

\bibitem[{\citenamefont{Niklasson et~al.}(2006)\citenamefont{Niklasson,
  Tymczak, and Challacombe}}]{ANiklasson06}
\bibinfo{author}{\bibfnamefont{A.~M.~N.} \bibnamefont{Niklasson}},
  \bibinfo{author}{\bibfnamefont{C.~J.} \bibnamefont{Tymczak}},
  \bibnamefont{and}
  \bibinfo{author}{\bibfnamefont{M.}~\bibnamefont{Challacombe}},
  \bibinfo{journal}{Phys. Rev. Lett.} \textbf{\bibinfo{volume}{97}},
  \bibinfo{pages}{123001} (\bibinfo{year}{2006}).

\bibitem[{\citenamefont{K\"{u}hne et~al.}(2006)\citenamefont{K\"{u}hne, Krack,
  Mohamed, and Parrinello}}]{TDKuhne07}
\bibinfo{author}{\bibfnamefont{T.~D.} \bibnamefont{K\"{u}hne}},
  \bibinfo{author}{\bibfnamefont{M.}~\bibnamefont{Krack}},
  \bibinfo{author}{\bibfnamefont{F.~R.} \bibnamefont{Mohamed}},
  \bibnamefont{and}
  \bibinfo{author}{\bibfnamefont{M.}~\bibnamefont{Parrinello}},
  \bibinfo{journal}{Phys. Rev. Lett.} \textbf{\bibinfo{volume}{98}},
  \bibinfo{pages}{066401} (\bibinfo{year}{2006}).

\bibitem[{\citenamefont{Niklasson}(2008)}]{ANiklasson08}
\bibinfo{author}{\bibfnamefont{A.~M.~N.} \bibnamefont{Niklasson}},
  \bibinfo{journal}{Phys. Rev. Lett.} \textbf{\bibinfo{volume}{100}},
  \bibinfo{pages}{123004} (\bibinfo{year}{2008}).

\bibitem[{\citenamefont{Steneteg et~al.}(2010)\citenamefont{Steneteg,
  Abrikosov, Weber, and Niklasson}}]{PSteneteg10}
\bibinfo{author}{\bibfnamefont{P.}~\bibnamefont{Steneteg}},
  \bibinfo{author}{\bibfnamefont{I.~A.} \bibnamefont{Abrikosov}},
  \bibinfo{author}{\bibfnamefont{V.}~\bibnamefont{Weber}}, \bibnamefont{and}
  \bibinfo{author}{\bibfnamefont{A.~M.~N.} \bibnamefont{Niklasson}},
  \bibinfo{journal}{Phys. Rev. B} \textbf{\bibinfo{volume}{82}},
  \bibinfo{pages}{075110} (\bibinfo{year}{2010}).

\bibitem[{\citenamefont{Niklasson and Cawkwell}(2012)\citenamefont{Niklasson
  and Cawkwell}}]{ANiklasson12}
\bibinfo{author}{\bibfnamefont{A.~M.~N.} \bibnamefont{Niklasson}},
  \bibnamefont{and}
  \bibinfo{author}{\bibfnamefont{M.~J.}~\bibnamefont{Cawkwell}},
  \bibinfo{journal}{Phys. Rev. B} \textbf{\bibinfo{volume}{86}},
  \bibinfo{pages}{174308} (\bibinfo{year}{2012}).

\bibitem[{\citenamefont{McWeeny}(1960)}]{RMcWeeny60}
\bibinfo{author}{\bibfnamefont{R.} \bibnamefont{McWeeny}},
  \bibinfo{journal}{Rev. Mod. Phys.} \textbf{\bibinfo{volume}{32}},
  \bibinfo{pages}{335} (\bibinfo{year}{1960}).

\bibitem[{\citenamefont{Cawkwell and Niklasson}(2012)}]{MJCawkwell12}
\bibinfo{author}{\bibfnamefont{M.~J.} \bibnamefont{Cawkwell}},
  \bibnamefont{and} \bibinfo{author}{\bibfnamefont{A.~M.~N.} \bibnamefont{Niklasson}},
  \bibinfo{journal}{J.\ Chem.\ Phys.\ } \textbf{\bibinfo{volume}{137}},
  \bibinfo{pages}{134105} (\bibinfo{year}{2012}).

\bibitem[{\citenamefont{UQuantChem}(2010)\citenamefont{Souvatzis}}]{UQuantChem}
\bibinfo{author}{\bibfnamefont{The UQuantChem code written by P.}~\bibnamefont{Souvatzis can be obtained from}},
  \bibinfo{pages}{ \text{http://www.anst.uu.se/pesou087/} } \text{UU-SITE/Webbplats\_2/UQUANTCHEM.html}, by using the password: "hylleraas" for decryption.

\bibitem[{\citenamefont{Niklasson et~al.}(2009)\citenamefont{Niklasson,
  Steneteg, Odell, Bock, Challacombe, Tymczak, Holmstr\"{o}m, Zheng, and
  Weber}}]{ANiklasson09}
\bibinfo{author}{\bibfnamefont{A.~M.~N.} \bibnamefont{Niklasson}},
  \bibinfo{author}{\bibfnamefont{P.}~\bibnamefont{Steneteg}},
  \bibinfo{author}{\bibfnamefont{A.}~\bibnamefont{Odell}},
  \bibinfo{author}{\bibfnamefont{N.}~\bibnamefont{Bock}},
  \bibinfo{author}{\bibfnamefont{M.}~\bibnamefont{Challacombe}},
  \bibinfo{author}{\bibfnamefont{C.~J.} \bibnamefont{Tymczak}},
  \bibinfo{author}{\bibfnamefont{E.}~\bibnamefont{Holmstr\"{o}m}},
  \bibinfo{author}{\bibfnamefont{G.}~\bibnamefont{Zheng}}, \bibnamefont{and}
  \bibinfo{author}{\bibfnamefont{V.}~\bibnamefont{Weber}}, \bibinfo{journal}{J.
  Chem. Phys.} \textbf{\bibinfo{volume}{130}}, \bibinfo{pages}{214109}
  (\bibinfo{year}{2009}).
  
\bibitem[{\citenamefont{Pulay}(1969)}]{PPulay69}
\bibinfo{author}{\bibfnamefont{P.} \bibnamefont{Pulay}},
  \bibinfo{journal}{Mol. Phys.} \textbf{\bibinfo{volume}{17}},
  \bibinfo{pages}{197} (\bibinfo{year}{1969}).

\bibitem[{\citenamefont{Roothaan}(1951)}]{Roothaan}
\bibinfo{author}{\bibfnamefont{C.~C.~J.} \bibnamefont{Roothaan}},
  \bibinfo{journal}{Rev. Mod. Phys.} \textbf{\bibinfo{volume}{23}},
  \bibinfo{pages}{69} (\bibinfo{year}{1951}).

\bibitem[{\citenamefont{Parr and Yang}(1989)}]{RParr89}
\bibinfo{author}{\bibfnamefont{R.~G.} \bibnamefont{Parr}} \bibnamefont{and}
  \bibinfo{author}{\bibfnamefont{W.}~\bibnamefont{Yang}},
  \emph{\bibinfo{title}{Density-functional theory of atoms and molecules}}
  (\bibinfo{publisher}{Oxford University Press}, \bibinfo{address}{Oxford},
  \bibinfo{year}{1989}).

\bibitem[{\citenamefont{Dreizler and Gross}(1990)}]{RMDreizler90}
\bibinfo{author}{\bibfnamefont{R.~M.} \bibnamefont{Dreizler}} \bibnamefont{and}
  \bibinfo{author}{\bibfnamefont{K.~U.} \bibnamefont{Gross}},
  \emph{\bibinfo{title}{Density-functional theory}}
  (\bibinfo{publisher}{Springer Verlag}, \bibinfo{address}{Berlin Heidelberg},
  \bibinfo{year}{1990}).

\bibitem[{\citenamefont{Becke}(1939)}]{ADBecke93}
\bibinfo{author}{\bibfnamefont{A.~D.} \bibnamefont{Becke}},
  \bibinfo{journal}{J.~Chem.~Phys.} \textbf{\bibinfo{volume}{98}},
  \bibinfo{pages}{1372} (\bibinfo{year}{1993}).


\bibitem[{\citenamefont{Ehrenfest}(1927)}]{PEhrenfest27}
\bibinfo{author}{\bibfnamefont{P.} \bibnamefont{Ehrenfest}},
  \bibinfo{journal}{Z. Phys.} \textbf{\bibinfo{volume}{45}},
  \bibinfo{pages}{455} (\bibinfo{year}{1927}).

\bibitem[{\citenamefont{Alonso}(2008)}]{JAlonso08}
\bibinfo{author}{\bibfnamefont{J.~L.} \bibnamefont{Alonso}},
\bibinfo{author}{\bibfnamefont{X.} \bibnamefont{Andrade}},
\bibinfo{author}{\bibfnamefont{P.} \bibnamefont{Echenique}},
\bibinfo{author}{\bibfnamefont{F.} \bibnamefont{Falceto}},
\bibinfo{author}{\bibfnamefont{D.} \bibnamefont{Prada-Garcia}},
\bibinfo{author}{\bibfnamefont{A.} \bibnamefont{Rubio}},
  \bibinfo{journal}{Phys. Rev. Lett.} \textbf{\bibinfo{volume}{101}},
  \bibinfo{pages}{096403} (\bibinfo{year}{2008}).

\bibitem[{\citenamefont{Jakowski}(2009)}]{JJakowski09}
\bibinfo{author}{\bibfnamefont{J.} \bibnamefont{Jakowski}}, \bibnamefont{and}
\bibinfo{author}{\bibfnamefont{K.} \bibnamefont{Morokuma}},
  \bibinfo{journal}{J. Chem. Phys.} \textbf{\bibinfo{volume}{130}},
  \bibinfo{pages}{224106} (\bibinfo{year}{2009}).



















\bibitem[{\citenamefont{Leimkuhler and Reich}(2004)}]{BLeimkuhler04}
\bibinfo{author}{\bibfnamefont{B.}~\bibnamefont{Leimkuhler}} \bibnamefont{and}
  \bibinfo{author}{\bibfnamefont{S.}~\bibnamefont{Reich}},
  \emph{\bibinfo{title}{Simulating Hamiltonian Dynamics}}
  (\bibinfo{publisher}{Cambridge University Press}, \bibinfo{year}{2004}).

\bibitem[{\citenamefont{Odell et~al.}(2009)\citenamefont{Odell, Delin,
  Johansson, Bock, Challacombe, and Niklasson}}]{AOdell09}
\bibinfo{author}{\bibfnamefont{A.}~\bibnamefont{Odell}},
  \bibinfo{author}{\bibfnamefont{A.}~\bibnamefont{Delin}},
  \bibinfo{author}{\bibfnamefont{B.}~\bibnamefont{Johansson}},
  \bibinfo{author}{\bibfnamefont{N.}~\bibnamefont{Bock}},
  \bibinfo{author}{\bibfnamefont{M.}~\bibnamefont{Challacombe}},
  \bibnamefont{and} \bibinfo{author}{\bibfnamefont{A.~M.~N.}
  \bibnamefont{Niklasson}}, \textbf{\bibinfo{volume}{131}},
  \bibinfo{pages}{244106} (\bibinfo{year}{2009}), \bibinfo{note}{J. Chem.
  Phys.}


%
%

%
%

\end{thebibliography}
\end{document}